\documentclass[preprintnumbers,article,amsmath,amssymb,floatfix,10pt,prd,onecolumn,superscriptaddress,nofootinbib]{revtex4-2}

\usepackage{titlesec}
\usepackage[toc]{appendix}

\titleformat*{\section}{\LARGE\bfseries}
\titleformat*{\subsection}{\Large\bfseries}
\titleformat*{\subsubsection}{\large\bfseries}
\titleformat*{\paragraph}{\large\bfseries}
\titleformat*{\subparagraph}{\large\bfseries}
\usepackage{bm}
\usepackage{amsfonts}
\usepackage{latexsym}
\usepackage[latin1]{inputenc}
\usepackage{graphicx}
\usepackage{amsmath}
\usepackage{palatino}
\usepackage{mathpazo}
\usepackage{textcomp}
\usepackage{float}
\usepackage{booktabs}
\usepackage{dcolumn}
\usepackage{ragged2e}
\usepackage{hyperref}
\hypersetup{colorlinks,citecolor=blue}
\hypersetup{colorlinks=true,linkcolor=red,filecolor=magenta,    urlcolor=blue}
\usepackage{amsthm}
\usepackage{xcolor}
\usepackage{orcidlink}
\usepackage{epsfig}
\usepackage[caption=false]{subfig}
\usepackage{commath}
\usepackage{cancel, ulem}\usepackage{adjustbox}

\usepackage{multirow}
\newcommand{\m}{\mathring}
\newcommand{\be}{\begin{equation}}
\newcommand{\ee}{\end{equation}}
\newcommand{\bea}{\begin{eqnarray}}
\newcommand{\eea}{\end{eqnarray}}
\newcommand{\eeas}{\end{eqnarray*}}
\newcommand{\beas}{\begin{eqnarray*}}

\def\jnl@style{\it}
\def\aaref@jnl#1{{\jnl@style#1}}

\def\aaref@jnl#1{{\jnl@style#1}}

\def\aj{\aaref@jnl{AJ}}                   
\def\apj{\aaref@jnl{ApJ}}                 
\def\apjl{\aaref@jnl{ApJ}}                
\def\apjs{\aaref@jnl{ApJS}}               
\def\apss{\aaref@jnl{Ap\&SS}}             
\def\aap{\aaref@jnl{A\&A}}                
\def\aapr{\aaref@jnl{A\&A~Rev.}}          
\def\aaps{\aaref@jnl{A\&AS}}              
\def\mnras{\aaref@jnl{Mon.~Not.~Roy.~Astron.~Soc.}}             
\def\prd{\aaref@jnl{Phys.~Rev.~D}}        
\def\prc{\aaref@jnl{Phys.~Rev.~C}}  
\def\prl{\aaref@jnl{Phys.~Rev.~Lett.}}    
\def\qjras{\aaref@jnl{QJRAS}}             
\def\skytel{\aaref@jnl{S\&T}}             
\def\ssr{\aaref@jnl{Space~Sci.~Rev.}}     
\def\zap{\aaref@jnl{ZAp}}                 
\def\nat{\aaref@jnl{Nature}}              
\def\aplett{\aaref@jnl{Astrophys.~Lett.}} 
\def\apspr{\aaref@jnl{Astrophys.~Space~Phys.~Res.}} 
\def\physrep{\aaref@jnl{Phys.~Rep.}}      
\def\physscr{\aaref@jnl{Phys.~Scr}}       
\def\commat{\aaref@jnl{Comm.~Math.~Phys.}}              
\def\science{\aaref@jnl{Science}}               
\def\cqg{\aaref@jnl{Classical Quant.~Grav.}}            
\def\jpcs{\aaref@jnl{JPCS}}                                     
\def\ijmpd{\aaref@jnl{Int.~J.~Mod.~Phys.~D}}                    
\def\grg{\aaref@jnl{Gen.~Relat.~Gravit.}}               
\def\rpp{\aaref@jnl{Rep.~Prog.~Phys.}}          
\def\npa{\aaref@jnl{Nucl.~Phys.~A}}        
\def\lrr{\aaref@jnl{Living Rev.~Rel.}}                   
\def\jcap{\aaref@jnl{J.~Cosmology Astropart.~Phys.}}    
\def\rmp{\aaref@jnl{Rev.~Mod.~Phys.}}   
\def\epjc{\aaref@jnl{Eur.~Phys.~J.~C}} 
\def\plb{\aaref@jnl{~Phy.~Lett.~B}} 
\def\mpla{\aaref@jnl{Mod.~Phy.~Lett.~A}} 
\def\arxiv{\aaref@jnl{arxiv.org}}

\allowdisplaybreaks[1]

\begin{document}
\title{The cosmological significance of boundary term in non-metricity gravity}
\author{Hamid Shabani\orcidlink{0000-0002-2309-3591}}
\email{h.shabani@phys.usb.ac.ir}
\affiliation{Physics Department, Faculty of Sciences, University of Sistan and Baluchestan, Zahedan, Iran}
\author{Avik De\orcidlink{0000-0001-6475-3085}}
\email{de.math@gmail.com}
\affiliation{Department of Mathematical and Actuarial Sciences, Universiti Tunku Abdul Rahman, Jalan Sungai Long,
43000 Cheras, Malaysia}
\author{Tee-How Loo\orcidlink{0000-0003-4099-9843}}
\email{looth@um.edu.my}
\affiliation{Institute of Mathematical Sciences, Faculty of Science, Universiti Malaya, 50603 Kuala Lumpur, Malaysia}


\footnotetext{The research was supported by the Ministry of Higher Education (MoHE), through the Fundamental Research Grant Scheme \ 
 (FRGS/1/2023/STG07/UM/02/3, project no.: FP074-2023).
 }

\begin{abstract}
Within the context of metric-affine gravity, we examine the significance of the boundary term in symmetric teleparallel gravity by employing the cosmological dynamical system analysis method. We focus on the novel gravity models characterized by the functions $f(Q,C)$, where $f$ is a smooth function of the non-metricity scalar $Q$ and the associated boundary term $C$. In a cosmological setting adopting three different classes of symmetric teleparallel affine connections, we investigate a model $f(Q,C)=Q^{s}+eC^{r}$, and some special cases of this model. We show that the boundary term which is added to the Einsteinian field equations (or equivalently to $f(Q)=Q$ ones) are capable of bringing forward solutions corresponding to the early accelerated expansion. This alludes the physics behind the boundary terms which usually are discarded in the most gravitational theories.    
\end{abstract}

\maketitle

\section{\textbf{Introduction}}
General Relativity (GR) is by far the most extensively validated theory of gravity. Nevertheless, the current concept of gravitational interaction is inadequate to explain phenomena at astrophysical and cosmic scales, including the clustering of objects and the accelerated expansion. So until a comprehensive quantum gravity theory is formulated, additional extensions or modification to GR is inevitable. The simplest modification of GR probably began with $f(\mathring{R})$ theory in which the Ricci scalar $\mathring{R}$ (corresponding to the torsion-free and metric-compatible Levi-Civita connection $\mathring{\Gamma}$) in Einstein-Hilbert action of GR was replaced by an arbitrary function $f(\mathring{R})$. There were several modified gravity theories originated from such curvature based geometry, the torsion-free and metric-compatibility quality of the geometry remained untouched. All such past attempts were made embracing GR in certain limits and allowing for more degrees of freedom, such that we can expand the scope of our analysis \cite{CANTATA:2021ktz,Boehmer:2021aji}.

There is an equally significant potential to develop a fundamental theory that addresses the limitations of GR, if one relaxes the torsion-free and/or metric-compatibility and employ a generic affine-connection instead of the Levi-Civita connection \cite{Boehmer:2023fyl,Iosifidis:2023pvz}. Within the realm of such metric-affine theories, symmetric  teleparallel gravity and its expansions are gaining significant importance in the ongoing discourse surrounding the development of a comprehensive theory of gravity. The term ``symmetric" here stands for the non-closedness of parallelograms generated by parallel transport of two vectors, i.e., zero torsion. Whereas the term ``teleparallel" has the implication that no change occured in parallel transporting a vector on a closed loop, i.e., zero curvature. This technique characterizes the geometrodynamical effects using the non-metricity nature of the affine connection with zero curvature and torsion. Ofcourse unlike the Levi-Civita connection based GR, the metric tensor and the affine connection here are treated as independent entities interacting through field equations. The most rigorously studied gravity theory in this class is the $f(Q)$ theory \cite{coincident, cosmology_Q, latetime, cosmography}, where $f$ is an arbitrary function of the non-metricity scalar $Q$ derived from the non-metricty character of the connection (systematically defined in the next section). This scalar $Q$ deviates from the Ricci scalar $\mathring R$ (derived from the unique Levi-Civita connection) by a boundary term denoted by $C$ as displayed in (\ref{aaboundary}). The $f(Q)$ theory has been introduced in the same route as $f(\mathring R)$ theory has been introduced to extend GR.  So, a linear function $f(Q)=\alpha Q+\beta$ eventually yields the same dynamics as GR and at Lagrangian level just differ by a boundary term \cite{trinity/boundary}. For this reason, this special linear case is also termed as symmetric teleparallel equivalent of GR (STEGR). 
Among other important works, The cosmological phase-space analysis of $f(Q)$ theory has been carried out in several instances which disclosed the interesting characteristics of this theory \cite{lu, shabani2023, Shabani:2023xfn, kyllep2022}. A thorough survey on $f(Q)$ theory of gravity and its cosmological and astrophysical applications can be found in \cite{lavinia/survey}.

Now, a major difference between $f(\mathring R)$ and $f(Q)$ theories lies in the fact that the latter is a second order theory like GR, whereas $f(\mathring R)$ theory gives us a fourth order field equation. One way to increase the order in symmetric teleparallel theories is to introduce higher order terms like $\Box Q, \Box^k Q$ in the Lagrangian \cite{saridakis/higher order}. However, a more natural approach is to incorporate the boundary term $C$ in the Lagrangian to develop $f(Q,C)$ theory as done in \cite{als/fQC,Capozziello:2023vne}. Take note that this $C$ is not the Gibbons-Hawking-York boundary like term that was required for a well-defined variational formulation for GR in spacetimes with boundaries \cite{ghy}. The resulting theory is popularly termed as the symmetric teleparallel equivalent of the $f(\mathring R)$ theory, since with a specific choice of function $f(Q,C)=f(Q+C)$ we can retrieve the $f(\mathring R)$ theory.

\section{Basic formulation of $f(Q,C)$ gravity}\label{formulation}
As discussed in the earlier section, the incompatibility of the affine connection with the metric, i.e., the non-vanishingness of the covariant derivative of the metric tensor is characterised by the non-metricity tensor
\begin{equation} \label{Q tensor}
Q_{\lambda\mu\nu} := \nabla_\lambda g_{\mu\nu}=\partial_\lambda g_{\mu\nu}-\Gamma^{\beta}_{\,\,\,\mu\lambda}g_{\beta\nu}-\Gamma^{\beta}_{\,\,\,\nu\lambda}g_{\beta\mu}\neq 0,
\end{equation}
and it demonstrates the gravitational interaction in such gravity theories solely.
We can always express the generic affine connection $\Gamma$ in terms of the Levi-Civita connection $\mathring \Gamma$ as
\begin{equation} \label{connc}
\Gamma^\lambda{}_{\mu\nu} := \mathring{\Gamma}^\lambda{}_{\mu\nu}+L^\lambda{}_{\mu\nu},
\end{equation}
where $L^\lambda{}_{\mu\nu}$ is the disformation tensor.
It follows that
\begin{equation} \label{L}
L^\lambda{}_{\mu\nu} = \frac{1}{2} (Q^\lambda{}_{\mu\nu} - Q_\mu{}^\lambda{}_\nu - Q_\nu{}^\lambda{}_\mu) \,.
\end{equation}
We can construct two different types of non-metricity vectors
\begin{equation*}
 Q_\mu := g^{\nu\lambda}Q_{\mu\nu\lambda} = Q_\mu{}^\nu{}_\nu \,, \qquad \tilde{Q}_\mu := g^{\nu\lambda}Q_{\nu\mu\lambda} = Q_{\nu\mu}{}^\nu \,.
\end{equation*}
Likewise, we write
\begin{align}
 L_\mu := L_\mu{}^\nu{}_\nu \,, \qquad 
 \tilde{L}_\mu := L_{\nu\mu}{}^\nu \,.   
\end{align}
The superpotential (or the non-metricity conjugate) tensor $P^\lambda{}_{\mu\nu}$ is given by
\begin{equation} \label{P}
P^\lambda{}_{\mu\nu} = 
\frac{1}{4} \left( -2 L^\lambda{}_{\mu\nu} + Q^\lambda g_{\mu\nu} - \tilde{Q}^\lambda g_{\mu\nu} -\delta^\lambda{}_{(\mu} Q_{\nu)} \right) \,.
\end{equation}
Finally, the non-metricity scalar $Q$ is defined as
\begin{equation} \label{Q}
Q=Q_{\alpha\beta\gamma}P^{\alpha\beta\gamma}\,.
\end{equation}
One can further obtain the following relations:
\begin{align}
\m R_{\mu\nu}+\m\nabla_\alpha L^\alpha{}_{\mu\nu}-\m\nabla_\nu\tilde L_\mu
+\tilde L_\alpha L^\alpha{}_{\mu\nu}-L_{\alpha\beta\nu}L^{\beta\alpha}{}_\mu=0\,,
\label{mRicci}\\
\m R+\m\nabla_\alpha(L^\alpha-\tilde L^\alpha)-Q=0\,. \label{mR}
\end{align}
As $Q^\alpha-\tilde Q^\alpha=L^\alpha-\tilde L^\alpha$, 
from the preceding relation, one also defines the boundary term as \label{aaboundary}
\begin{align}
C=\m{R}-Q&=-\m\nabla_\alpha(Q^\alpha-\tilde Q^\alpha)
=-\frac1{\sqrt{-g}}\partial_\alpha\left[\sqrt{-g}(Q^\alpha-\tilde Q^\alpha)\right].
\end{align}
In view of the the above discussion, $f(Q,C)$ gravity is introduced  incorporating both $f(\m R)$ and $f(Q)$ theories into a more general framework. 
The action is defined by 
\begin{equation}
S=\int \left[ \frac{1}{2\kappa }f(Q,C)+\mathcal{L}_{M}\right] \sqrt{-g}%
\,d^{4}x\,,
\label{eqn:action-fQC}
\end{equation}
where $f$ is a function on both $Q$ and $C$; and $\mathcal L_M$ is a matter Lagrangian. The metric field equation is given by
\begin{align}\label{eqn:FE1}
\kappa T_{\mu\nu}
=&-\frac f2g_{\mu\nu}+2P^\lambda{}_{\mu\nu}\nabla_\lambda(f_Q-f_C)
  +\left(\m G_{\mu\nu}+\frac Q2g_{\mu\nu}\right)f_Q
  \notag\\&+\left(\frac C2g_{\mu\nu}-\m\nabla_{\mu}\m\nabla_{\nu}
  +g_{\mu\nu}\m\nabla^\alpha\m\nabla_\alpha \right)f_C\,.
\end{align}
It is important to keep in mind that, unlike GR, in the teleparallel theory the affine connection is indepedent of the metric tensor, and both act as dynamic variables. Therefore, by taking variation of the action with respect to the affine connection, we obtain the connection field equation
\begin{align}\label{eqn:FE2-invar}
(\nabla_\mu-\tilde L_\mu)(\nabla_\nu-\tilde L_\nu)
\left[4(f_Q-f_C)P^{\mu\nu}{}_\lambda\right]=0\,,
\end{align}
in the absence of the hypermomentum tensor.

The present paper attempts to assess the cosmological significance of 
$f(Q,C)$ gravity. 
To achieve this goal, we consider a spatially 
flat Friedmann-Robertson-Walker (FLRW) geometry 
\begin{equation}
ds^{2}=-dt^{2}+a^{2}(t)[dx^{2}+dy^{2}+dz^{2}],  \label{3a}
\end{equation}%
having a homogeneous and isotropic matter distribution,
where $a(t)$ is said to be the scale factor of the Universe, and its first time derivative is given by the Hubble parameter $H(t)=\dot{a}(t)/a(t)$. Here $\dot{()}$ indicates a derivative with respect to cosmic time $t$. 

The FLRW metric (\ref{3a}) exhibits cosmic symmetry, specifically homogeneity and isotropy, which can be described by spatial rotational and translational transformations. A symmetric teleparallel affine connection $\Gamma$ is an affine connection that is both torsion-free and curvature-free. It also possesses both spherical and translational symmetries, implying vanishing of the Lie derivative of the affine connection in terms of the Killing vectors. There exist three classes of such affine connections characterized by a temporal parameter $\gamma(t)$, as outlined in the references \cite{FLRW/connection, ad/bianchi, fQec1}. Let us denote these classes by Connection type $I$, $II$ and $III$. For specific cosmological applications, it is necessary to analyze each class of these symmetric teleparallel connections and establish the associated equations of motion for $f(Q,C)$ gravity, which were derived in \cite{als/fQC}. We analyse each phenomenon in the following sections.

\section{Connection type $I$}\label{coin}
In this section we briefly describe the cosmological behavior corresponding to the first class of connection with $\gamma(t)=0$. As discussed in the literature \cite{FLRW/connection, ad/bianchi}, under a gauge transformation to the Cartesian coordinates the connection components all vanish, $\Gamma_{\mu\nu}^{\alpha}=0$. Under this coincident gauge, $f(Q,C)$ theory gives the same field equations as those of $f(T,B)$ gravity \cite{teleparallel}\footnote{Here, we have $T=6H^{2}=-Q$ and $B=6(3H^{2}+\dot{H})$.}. Recently $f(T,B)$ gravity has been studied in~\cite{franco2020} and it has been asserted that these types of model include a solution describing a dark matter dominated epoch\footnote{Actually, the authors of~\cite{franco2020} getting some inconsistent equations (eqs. 5-10) have shown that models with $f(T,B)=f_{0}B^{k}T^{m}$ contains a solution which indicates the dark matter dominated era. However, the present authors believe that their results are uncertain.}. In this regard we present here some comments to complete our study. The Friedmann-like equations in this case are given by \cite{als/fQC}

\begin{align}
&\kappa\rho_{m}=6H^{2}f_{Q}+\frac{f}{2}-Cf_{C}+3H\dot{f}_{C},\label{coin.1}\\
&\kappa p_{m}=-6H^{2}f_{Q}-\frac{f}{2}+Cf_{C}-2\dot{H}f_{Q}-2H\dot{f}_{Q}-\ddot{f_{C}}\label{coin.2},
\end{align}
where we eliminate the argument $(Q,C)$ for the sake of abbreviation.

\subsection{$f(Q,C)=Q+dC^{2}$}\label{coins1}
As can be seen, all terms in eqs.~(\ref{coin.1}) and (\ref{coin.2}) depend on the function $f(Q,C)$ and its derivatives. To 
include the GR solutions in addition to the early or late time effects, one may tempt to choose those functions which give the GR equations of motion in case of vanishing extra terms. A choice can be the function $f(Q,C)=Q+dC^{v}$ for constants $d$ and $v$. Setting $v=2$ for a simple presentation, one gets

\begin{align}
&\kappa\rho_{m}=3H^{2}-d\left(\frac{C^{2}}{2}-6H\dot{C}\right),\label{coin.3}\\
&\kappa p_{m}=-3H^{2}-2\dot{H}+d \left(\frac{C^{2}}{2}-2\ddot{C}\right).\label{coin.4}
\end{align}

We see that eqs.(\ref{coin.3}) and (\ref{coin.4}) restore the standard GR equations for matter dominated era when $d=0$. Assuming the pressureless perfect fluid fills the Universe one obtains the following dimensionless equations of motion.

\begin{align}
&\Omega_{de}=d(x_1-2x_2),\label{coin.55}\\
&\Omega_{m}=1-\Omega_{de},\label{coin.5}\\
&\frac{dx_{1}}{dN}=2(x_3+3)x_2-2x_1x_3,\label{coin.6}\\
&\frac{dx_{2}}{dN}=\frac{1}{d}\left[-\frac{3}{2}(1-dx_1)-x_3\right]-x_2x_3,\label{coin.7}\\
&\frac{dx_{3}}{dN}=(3+x_3)\left[\frac{x_2}{x_1}(3+x_3)-2x_3\right],\label{coin.8}
\end{align}
where $N=\ln a$ and we have defined the following dimensionless variables

\begin{align}\label{coin.9}
\Omega_{m}=\frac{\kappa \rho_m}{3H^2},~~~~~x_1=\frac{C^2}{6H^2},~~~~~x_2=\frac{\dot{C}}{H},~~~~~x_3=\frac{\dot{H}}{H^2}.
\end{align}

\begin{figure}[h!]
\centering
\includegraphics[width=.45\textwidth]{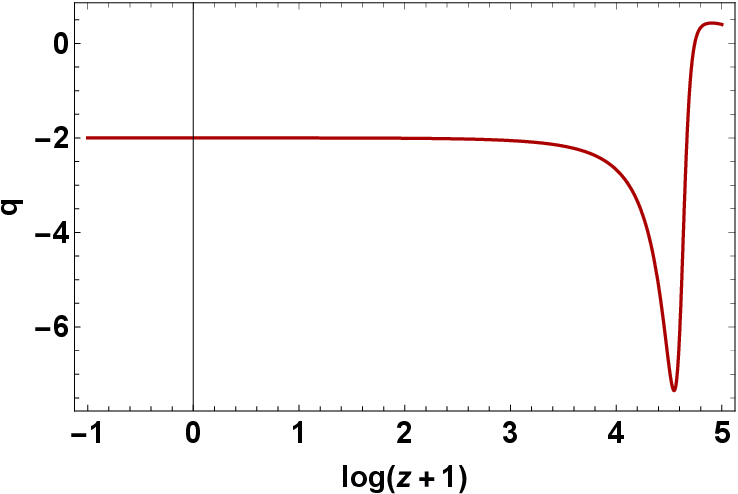}\hspace{2mm}
\includegraphics[width=.45\textwidth]{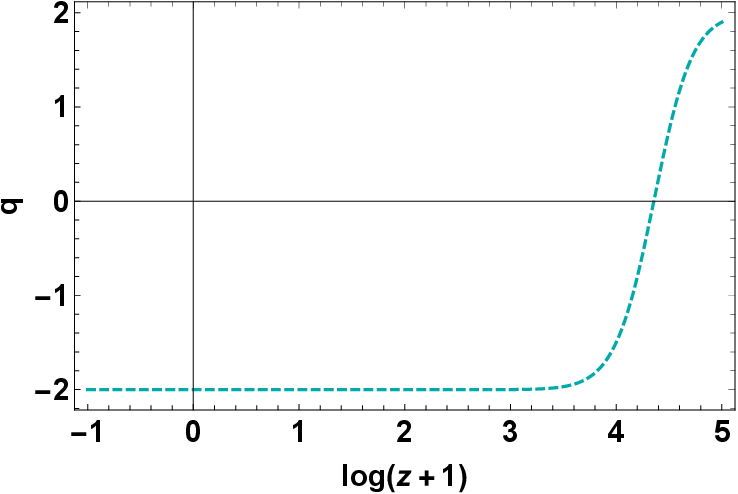}\vspace{2mm}
\includegraphics[width=.45\textwidth]{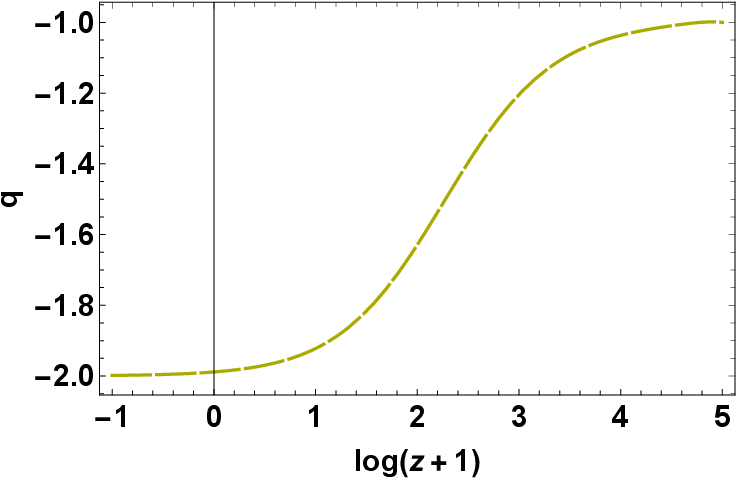}\hspace{2mm}
\includegraphics[width=.45\textwidth]{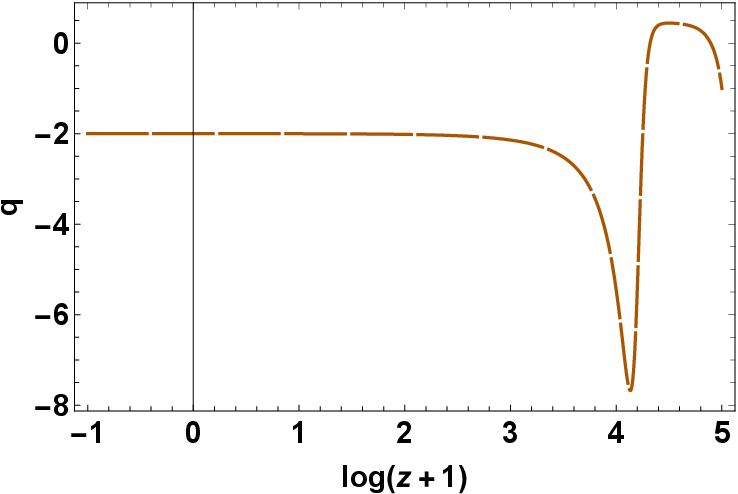}\vspace{2mm}
\includegraphics[width=.45\textwidth]{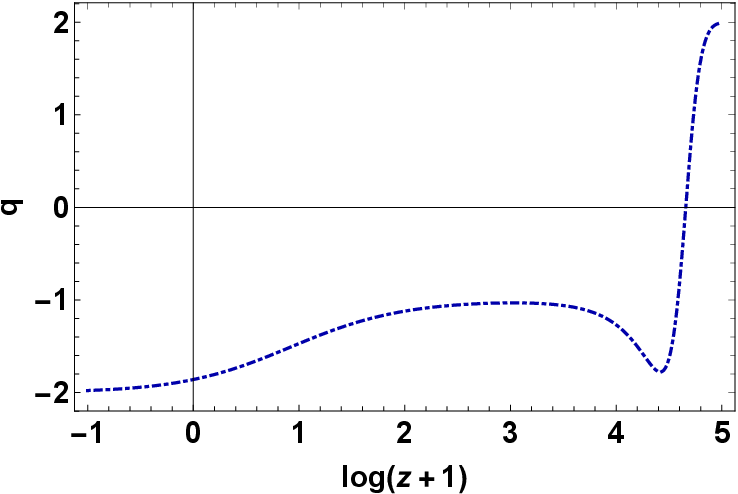}\
\caption{The behavior of the deceleration parameter predicated by the model $f(Q,C)=Q+dC^{2}$. $\alpha=-10$ and initial values $x_{1i}=0.001, x_{2i}=-0.002, x_{3i}=-1.40$ for the upper left panel, $\alpha=4, x_{1i}=-0.625, x_{2i}=1.25, x_{3i}=-2.9$ for the upper right panel, $\alpha=4, x_{1i}=0.255, x_{2i}=-0.001,x_{3i}=2\times 10^{-5}$ in the case of the middle left one, $\alpha=-6, x_{1i}=0.0138164, x_{2i}=-0.0276328, x_{3i}=1\times 10^{-5}$ in the case of the middle right panel and $\alpha=-10, x_{1i}=-2.98\times10^{-6}, x_{2i}=5.96\times10^{-6}, x_{3i}=-2.99$ for the lower diagram have been used.}\label{coin1}
\end{figure}

Now, the main problem appears. As Table~\ref{tab0} shows there are four critical points; $P_{ph}$ is the only attractor solution which corresponds to a phantom dark energy era with $q=-2$, a dark matter dominated solution, $P_m$, a solution which denotes a stiff fluid dominated phase $P_{st}$ and $P_{ds}$ display a de Sitter phase. Only the de Sitter fixed point is an exact solution of the system~(\ref{coin.6})-(\ref{coin.8}) and the others are accounted for approximate ones (see the third and the last column of Table~\ref{tab0}). These critical points with the mentioned stability properties make possible different transitions in phase space. Fig.~\ref{coin1} does sketch some of these behaviors. For example, the solid curve implies a transition from the matter dominated phase to the phantom era. Its right panel shows a stiff fluid $\to$ phantom evolution. The middle left panel displays a de Sitter phase which directly connect to the phantom phase. An exciting case is a de Sitter $\to$ dark matter $\to$ phantom transition (it has been drawn in the middle right panel) which includes an early and late expansion eras. Furthermore, there is a stiff fluid dominated $\to$ de Sitter $\to$ phantom transition (the lower panel).

\begin{table}[h!]
\centering
\caption{The fixed points solutions of $f(Q)=Q+d C^{2}$ gravity. The primes over the variables $x_i$ denote $d/dN$.}
\begin{adjustbox}{width=1\textwidth}
\begin{tabular}{l @{\hskip 0.1in}l @{\hskip 0.1in} l@{\hskip 0.1in} l @{\hskip 0.1in}l @{\hskip 0.1in}@{\hskip 0.1in}ll}\hline\hline

Fixed point     &$(x_1,x_2,x_3,)$      & $(x'_1,x'_2,x'_3)$     &Eigenvalues  &$\Omega_m$      &$q$&type\\[0.5 ex]
\hline
$P_m$       & $\left(x_1,-2x_1,-\frac{3}{2}\right)$ 
            & $(-3x_1,-\frac{3}{2}x_1,0)$   
            & $\left[-3,\frac{3}{2 \sqrt{-\alpha x_1 } }+\frac{3}{4},
                    -\frac{3}{2 \sqrt{-\alpha x_1 }}+\frac{3}{4}\right]$
            & $1-5 \alpha  x_1$
            & $\frac{1}{2}$
            & $|x_1|\ll 1$\\[0.75 ex]
$P_{ds}$    & $\left(\frac{1}{d},0,0\right)$        
            & $(0,0,0)$
            & $\left[-3,\frac{3}{2} \left(\sqrt{5}-1\right),
                -\frac{3}{2}\left(\sqrt{5}+1\right)\right]$
            & $-1$ 
            & $-1$
            & g\\[0.75 ex]
$P_{st}$    & $\left(x_1,-2x_1,-3\right)$           
            & $[6x_1,\frac{3}{2\alpha} \left(1-3 x_1\alpha\right),0]$
            & $\left[3,6,6\right]$
            & $1 - 5 \alpha x_1 $
            & $2$
            & $|x_1\alpha|\ll 1$, \\[0.75 ex]
            &&&&&&  $\alpha\gg1$\\[0.75 ex]
$P_{ph}$    & $\left(x_1,x_2,1\right)$              
            & $[-2x_1,\frac{3}{2}x_1,-2 (x_3+3) x_3]$
            & $\left[-2,0,-10\right]$
            & $-\alpha x_1$
            & $-2$
            & $x_1\gg1$   \\[0.75 ex]
\hline\hline
\end{tabular}
\end{adjustbox}
\label{tab0}
\end{table}



\subsection{$f(Q,C)=Q^{s}+eC^{r}$}\label{coins2}
One may also tempt to choose the generalized form of the function analysed in subsection~\ref{coins1}. Since, we work on $f(Q,C)=Q^{s}+eC^{r}$ theories and gain the corresponding field equations. Using the following variables

\begin{align}\label{coin.10}
\Omega_{m}=\frac{\kappa \rho_m}{3H^2},~~
y_1=(1-2s)\frac{Q^s}{6H^2},~~
y_2=e(1-r)\frac{C^r}{6H^2},~~
y_3=er(r-1)\frac{C^{r-2}\dot{C}}{H},~~
y_4=\frac{\dot{H}}{H^2},
\end{align}

the field equations can be obtained as

\begin{align}
&\Omega_{m}=y_{1}+y_{2}+y_{3},\label{coin.11}\\
&\frac{dy_{1}}{dN}=2 (e - 1) y_{1} y_{4},\label{coin.12}\\
&\frac{dy_{2}}{dN}=-r y_{3} (3 + y_{4}) - 2y_{2} y_{4},\label{coin.13}\\
&\frac{dy_{3}}{dN}=\frac{2 s (3-2 s) }{1-2 s}y_{1} y_{4}-3 (y_{1}+y_{2})-y_{3} y_{4},\label{coin.14}\\
&\frac{dy_{4}}{dN}=-(y_{4}+3) \left(\frac{y_{3} (y_{4}+3)}{y_{2}}+2y_{4}\right).\label{coin.15}
\end{align}

\begin{figure}[h!]
\centering
\includegraphics[width=.45\textwidth]{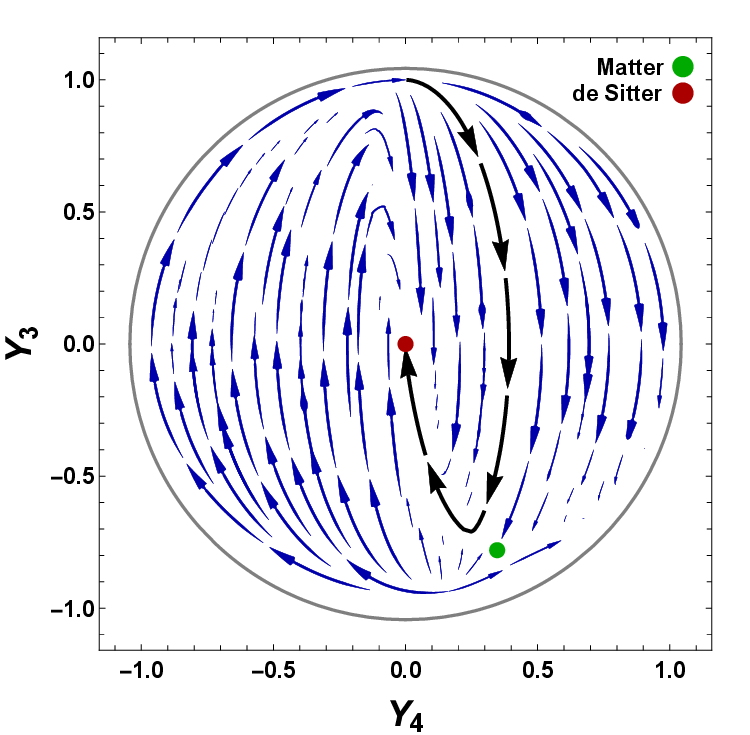}\hspace{2mm}
\includegraphics[width=.45\textwidth]{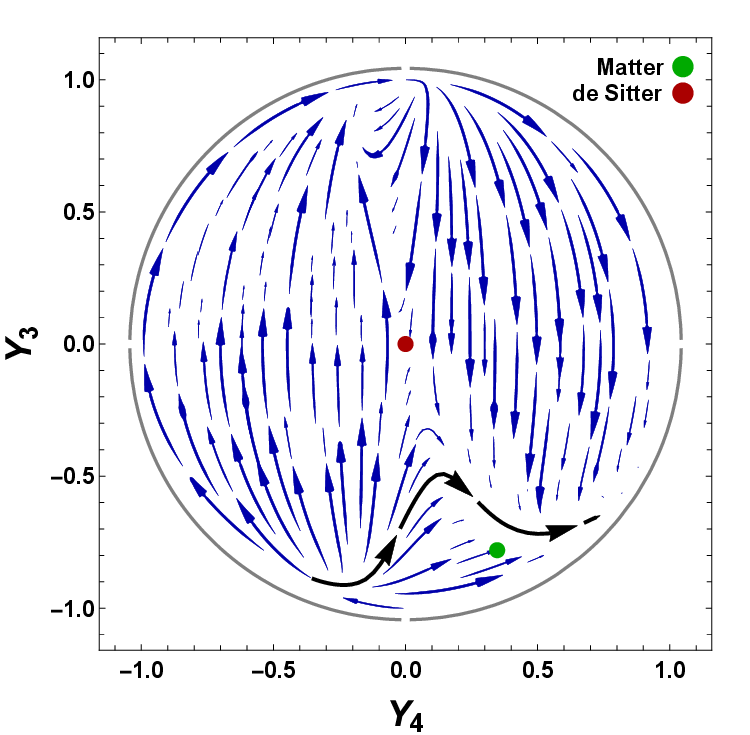}\vspace{2mm}
\includegraphics[width=.45\textwidth]{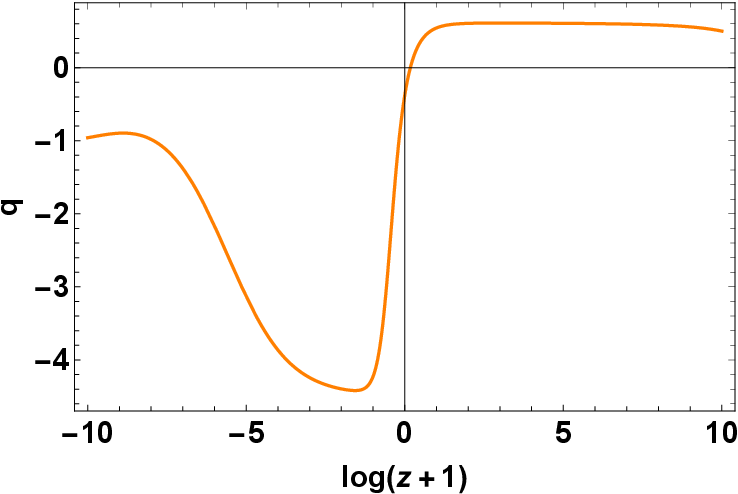}\hspace{2mm}
\includegraphics[width=.45\textwidth]{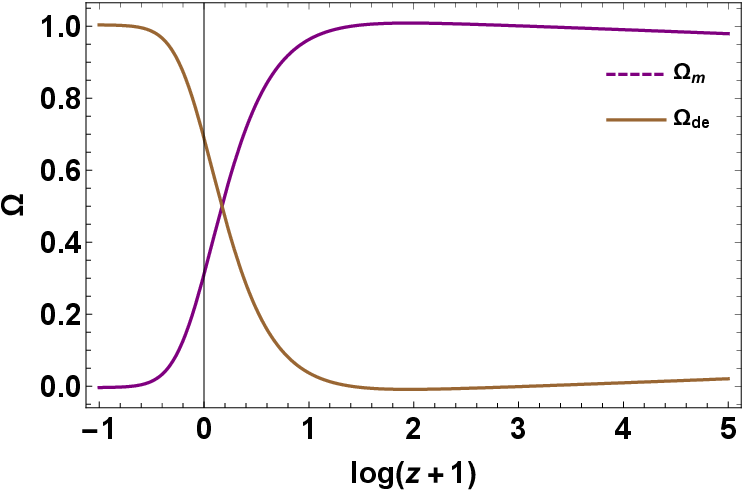}
\caption{The related plots of the models $f(Q,C)=Q^{s}+e C^{r}$in the non-coincident gauge. The left phase space has been drawn for $s=1.08$ and the right one depicted for $s=3$. The lower diagrams describe a transition form a dark matter dominated phase to a transient phantom phase followed by a de Sitter era. We have set the initial values $y_{1i}=-0.500, y_{2i}=0.323, y_{3i}=0.660, y_{4i}=-1.5, r=1$ and $s=1.05$ to plot the lower diagrams.}\label{noncoinsr}
\end{figure}

The system of equations (\ref{coin.12})-(\ref{coin.15}) has four fixed point solutions whose details are explained in Table~\ref{tab2}. In the case of $f(Q,C)=Q^{s}+eC^{r}$ the matter dominated era appears only for $r=1$ for arbitrary values of $e$ and $s$. Also, there is an unstable phantom era which is followed by a stable de Sitter epoch for $1.00<s\lesssim1.16$. The upper left panel in Fig.~\ref{noncoinsr} represent an example in which the de Sitter solution is stable and the upper right one describes a phase transition to an unstable de Sitter era. Also, a transition of type dark matter $\to$ phantom $\to$ de Sitter is illustrated in lower panels Fig.~\ref{noncoinsr}.

\begin{table}[h!]
\centering
\caption{The critical point obtained for $f(Q,C)=Q^{s}+e C^{r}$.}
\begin{tabular}{l @{\hskip 0.1in} l@{\hskip 0.1in} l @{\hskip 0.1in}l @{\hskip 0.1in}l}\hline\hline

Fixed point     &$(y_1,y_2,y_3,y_4)$           &Eigenvalues  &$\Omega_m$      &$q$\\[0.5 ex]
\hline\vspace{2mm}
$P_m$&$\left(0,\frac{1}{3},\frac{2}{3},-\frac{3}{2}\right)$&$\left[0,3(1- s),\frac{3}{4} \left(3+i \sqrt{15}\right),\frac{3}{4} \left(3-i \sqrt{15}\right)\right]$&$1$&$\frac{1}{2}$\\[0.75 ex]
$P_{ds}$&$\left(x_1,-x_1,0,0\right)$&stable $\forall 1.00<s<1.16$&$0$&$-1$\\[0.75 ex]
$P_{st}$&$\left(0,0,0,-3\right)$&$\left[3, 6, \infty, 6 (1 - s)\right]$&$0$&$2$\\[0.75 ex]
$P_{ph}$&$\left(0,x_2,-x_2,3\right)$&unstable $\forall s$&$0$&$-4$\\[0.75 ex]
\hline\hline
\end{tabular}
\label{tab2}
\end{table}

In the next section, we consider $f(Q,C)$ gravity with non-trivial affine connection involved in classes $\Gamma_{II}$ and $\Gamma_{III}$.


\section{$f(Q,C)$ gravity in non-trivial connections}\label{noncoin1}

In the present section we explore the non-vanishing connections. The corresponding dimensionless system of equations are introduced and their cosmological behavior are discussed. We use the function $f(Q,C)=Q^{n}+\alpha C^{m}$. Note that both connections leads to the same non-metricity scalar $Q=-6H^{2}$ for a particular $\gamma(t)$ as discussed in \cite{als/fQC}.

\section{Connections type $II$}\label{noncoin1.1}
Equations of motion for this class of connections become \cite{als/fQC}
 
\begin{align}
&\kappa \rho_m=3H^2+\frac{f}{2}+3H^{2}f_{Q}-C\frac{f_{C}}{2}+\frac{3\gamma}{2a^{3}}\dot{f}_{Q}+\left(3H-\frac{3\gamma}{2a^{3}}\right)\dot{f}_{C},\label{noncoin.1}\\
&\kappa p_m=-2\dot{H}-3H^{2}-\frac{f}{2}-3H^{2}f_{Q}+C\frac{f_{C}}{2}+\left(\frac{3\gamma}{2a^{3}}-2H\right)\dot{f}_{Q}-\frac{3\gamma}{2a^{3}}\dot{f}_{C}-\ddot{f}_{C},\label{noncoin.2}\\
&\dot{\rho}_{m}+3H(\rho_{m}+p_{m})=\frac{3\gamma}{2\kappa a^{3}}\left[3H\dot{f}_{Q}+\ddot{f}_{Q}-(3H\dot{f}_{C}+\ddot{f}_{C})\right].\label{noncoin.3}
\end{align}

The conservation of EMT assumption forces the right hand side of eq.~(\ref{noncoin.3}) to be zero. One option which is consistent with the function $f(Q,C)=Q^{n}+\alpha C^{m}$ reads

\begin{align}
&3H\dot{f}_{Q}+\ddot{f}_{Q}=0,\label{noncoin.4}\\
&3H\dot{f}_{C}+\ddot{f}_{C}=0.\label{noncoin.5}
\end{align}

We utilize the constraints (\ref{noncoin.4})-(\ref{noncoin.5}) to eliminated some variables of the corresponding phase space of system (\ref{noncoin.1})-(\ref{noncoin.3}). We consider the following dimensionless variables 

\begin{align}\label{noncoin.6}
\Omega_{m}=\frac{\kappa \rho_m}{3H^2},~~~~~x_1=\frac{Q^n}{6H^2},~~~~~x_2=\frac{C^m}{6H^2},~~~~~x_3=\frac{\dot{H}}{H^2},~~~~~x_4=\frac{1}{Ha^{3}},~~~~~x_5=\frac{H\dot{C}}{C^{2}}.
\end{align}

Not all of the variables~(\ref{noncoin.6}) are independent. In fact the equations (\ref{noncoin.4})-(\ref{noncoin.5}) give

\begin{align}
&x_{5}=\frac{x_3 \Big[(3-2 n) x_3+3\Big]}{6 \left(x_3+3\right)^2},\label{noncoin.7}\\
&x_{2}=\frac{2 \left(x_3+3\right)^2 \Big\{(n-1) x_1 \big[n x_3 \left(3 \gamma  x_4-4\right)-3\big]+2 x_3+3\Big\}}{9 \alpha  \gamma  m(m-1) x_3 x_4 \big[(2 n-3) x_3-3\big]+6 \alpha  (m-1) \left(x_3+3\right)^2+6 x_3 \big[(3-2 n) x_3+3\big]}.\label{noncoin.8}
\end{align}

Therefore, the dynamical system equivalent of the system (\ref{noncoin.9})-(\ref{noncoin.13}) keeping the variables $(x_{1},x_{3},x_{4})$ as independent ones, is obtained as

\begin{align}
&\Omega_{m}=1-\Omega_{de},\label{noncoin.9}\\
&\Omega_{de}=\alpha  (m-1) x_2 \big[3 m x_5 \left(\gamma  x_4-2\right)+1\big]+(n-1) x_1 \big[\gamma  n x_3 x_4+1\big],\label{noncoin.10}\\
&\frac{dx_{1}}{dN}=2 (n-1) x_1 x_3,\label{noncoin.11}\\
&\frac{dx_{3}}{dN}=2 \left(x_3+3\right) \big[3 \left(x_3+3\right) x_5-x_3\big],\label{noncoin.12}\\
&\frac{dx_{4}}{dN}=-x_4 (3 + x_3).\label{noncoin.13}
\end{align}

The system (\ref{noncoin.9})-(\ref{noncoin.13}) admits two critical points; de Sitter solution $p_{ds}$ which is an attractor independent on the phase space constants $\alpha, \gamma, m, n $ as well as a saddle one, i.e., $p_m$, whose physical properties depend on the model constants. Table~\ref{tab1} shows these solutions. As can be seen, these solutions do not contain a solution corresponding to the matter dominated era. Nevertheless, $P_m$ indicates the aspects a mater dominated fixed point for $n=3/2,m=5/3$, that is we have $\Omega_m=1$ and $w_{eff}=-1-2\dot{H}/3H^{2}=0$ in this case. 


\begin{table}[h!]
\centering
\caption{The fixed points solutions for models which are studied in Sect.~\ref{noncoin1}.}
\begin{tabular}{l @{\hskip 0.1in} l@{\hskip 0.1in} l @{\hskip 0.1in}l @{\hskip 0.1in}l}\hline\hline

Fixed point     &Coordinates $(x_1,x_3,x_4,)$           &Eigenvalues  &$\Omega_m$      &$w_{eff}$\\[0.5 ex]
\hline
&&Connection type II\\
\hline\vspace{2mm}
$P_m$&$\left(0,\frac{3}{1-2 n},0\right)$&$\left[\frac{3}{2 n-1}-3,\frac{3}{2 n-1}-3,3\right]$&$\frac{-\alpha  (m-1) \big[2 (m-1) n-3 m+2\big]-2 n+1}{\alpha  (m-1) (n-1) (2 n-1)-2 n+1}$&$\frac{2}{2 n-1}-1$\\[0.75 ex]
$P_{ds}$&$\left(x_1,0,0\right)$&$[0,-3,-3]$&$0$&$-1$\\[0.75 ex]
\hline
&&Connection type III\\
\hline\vspace{2mm}
$P_m$&$\left(0,\frac{1}{2 i+1},0\right)$&$\left[-1,1-\frac{3}{2 i+1},-\left(\frac{1}{2 i1}+3\right)\right]$&$\frac{\beta  (j-1) \big[6 i (j-1)+5 j-4\big]-2 i-1}{3 \beta  \big[ (6 i+7)+2\big] (j-1)-2 i-1}$&$-\frac{2}{6 i+3}-1$\\[0.75 ex]
$P_{ds}$&$\left(x_1,0,0\right)$&$\left[0,-3,1\right]$&$0$&$-1$\\[0.75 ex]
\hline\hline
\end{tabular}
\label{tab1}
\end{table}

\begin{figure}[h!]
\centering
\includegraphics[width=.45\textwidth]{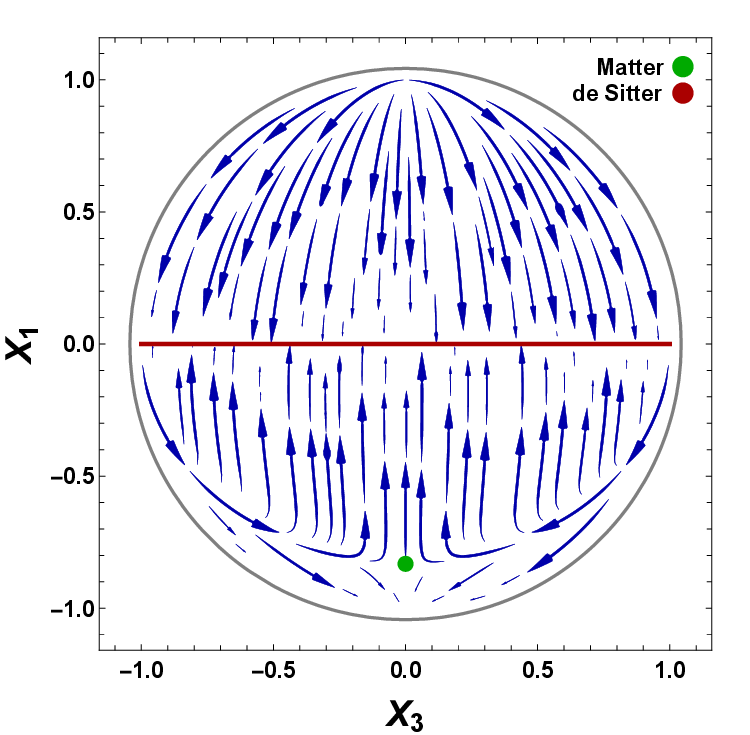}\hspace{2mm}
\includegraphics[width=.45\textwidth]{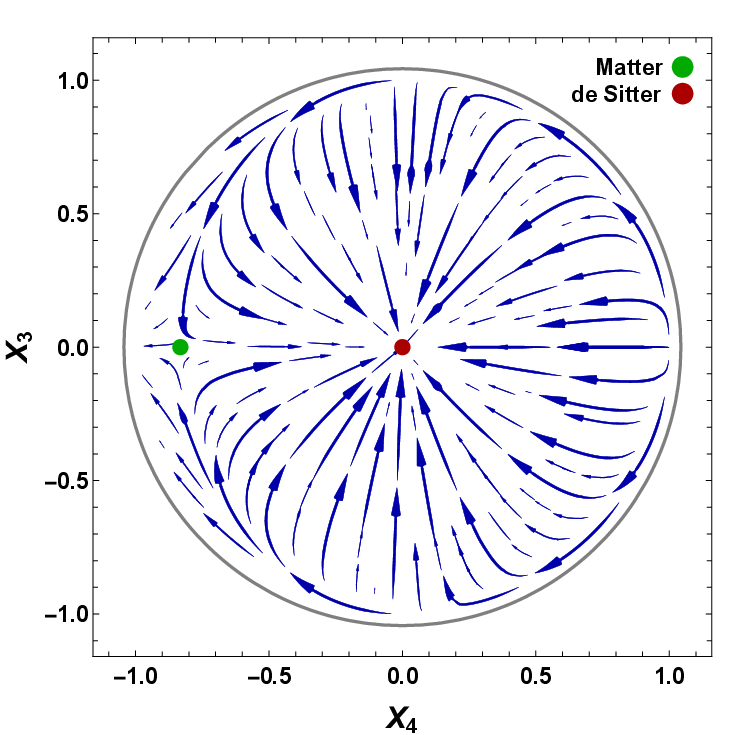}
\caption{Phase space diagrams for the models $f(Q,C)=Q^{\frac{3}{2}}+\alpha C^{\frac{5}{3}}$ when connection type $\Gamma_II$ has been assumed.}\label{noncoin11}
\end{figure}

\begin{figure}[h!]
\centering
\includegraphics[width=.45\textwidth]{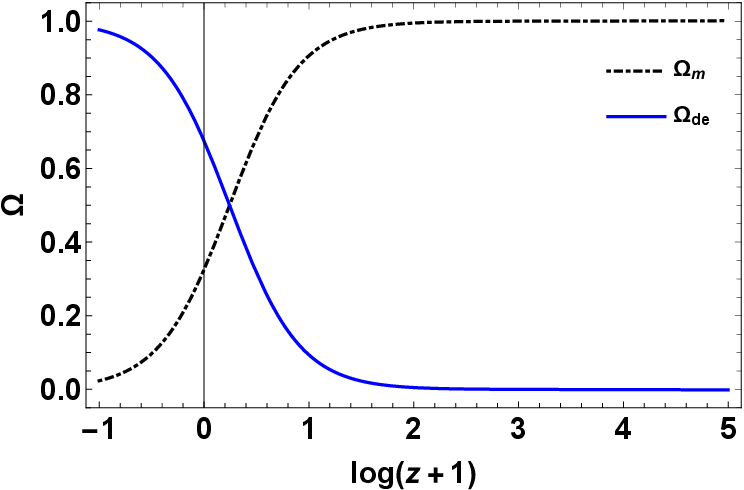}\hspace{2mm}
\includegraphics[width=.45\textwidth]{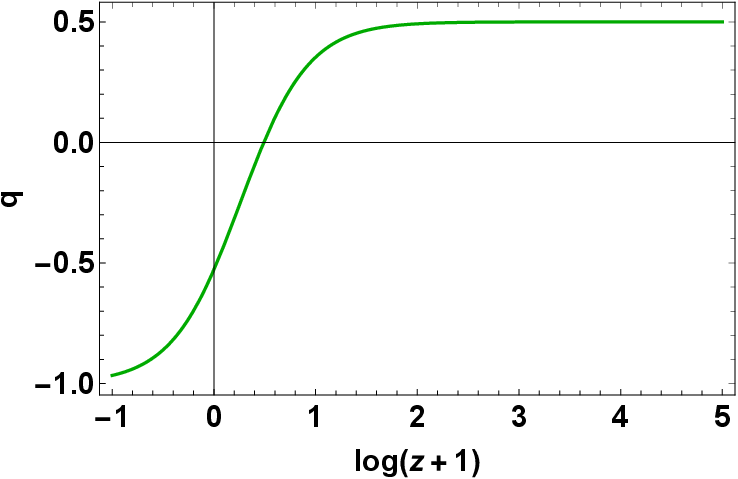}
\caption{Different cosmological parameters in $f(Q,C)=Q^{\frac{3}{2}}+\alpha C^{\frac{5}{3}}$ gravity. Left panel: the evolution of the matter density parameter, the black dashed curve, and the dark energy matter density, the blue solid one. Right panel: the evolution of the deceleration parameter. The initial values $x_{1i}=-1\times 10^{-3}$, $x_{3i}=-1.49$, $x_{4i}=-1\times 10^{-4}$ for $\alpha=-0.85$ and $\gamma=2.5$ have been used.}\label{noncoin12}
\end{figure}

Note that as Table~\ref{tab1} shows the critical point $P_m$ is always a saddle point while $p_{ds}$ behaves as an attractor one, independent on the free constants of model. 
In Fig.~\ref{noncoin11} we have illustrated the phase space trajectories in $(x_1,x_3)$ plane for $x_4=0$ and also $(x_3,x_4)$ plane for $x_1=0$. 
Also, in Fig.~\ref{noncoin12} the behavior of the matter density and the deceleration parameters have been demonstrated for $n=3/2,m=5/3$. 
As Fig.~\ref{noncoin11} and Fig.~\ref{noncoin12} 
display a desirable transition from the 
dark matter dominated era to the 
de Sitter one takes place. \\ 


The behavior of the Hubble parameter can also be derived using eqs.~(\ref{noncoin.1})-(\ref{noncoin.3}). Since, only two of these equations are independent, one can use any two of them. On the other hand, these equations are highly complicated  in the complete form, i.e., when a $f(Q,C)$ function is selected, because of many terms including different order of time differentiation of the Hubble parameter (up to three) and their multiplications. Particularly, when they are rewritten in terms of redshift the issue persists with a greater intricacy. Thus, we only discuss a $f(Q)$ function and leave a complete investigation, including $f(Q,C)$ functions, to an independent work. Besides, to avoid any mathematical rick we also leave the conservation of the matter energy density $\rho_m$ assumption. With the mentioned assumptions and considering $f(Q)=\eta Q^{3/2}$\footnote{In our previous paper we have showed that $f(Q)=\eta Q^{n}$ gravities admit a matter dominated followed by a de Sitter era for arbitrary values of $n$~\cite{shabani2023}.} eqs.~(\ref{noncoin.2})-(\ref{noncoin.3}) read
\begin{align}
&4(3 H^2+2H')+\eta\left\{\left[3 \sqrt{6}(6 H^2+1)\right] H'+\frac{3 \sqrt{6} \left(4 a^3 H-3 \gamma \right)}{a^3}H''\right\} =0,\label{noncoinH1}\\
&\dot{\rho}+3H\rho-\frac{9}{2} \sqrt{\frac{3}{2}} \gamma\eta\frac{ H''+3 H H'}{a^3}=0.\label{noncoinH2}
\end{align}

\begin{figure}[h!]
\centering
\includegraphics[width=.45\textwidth]{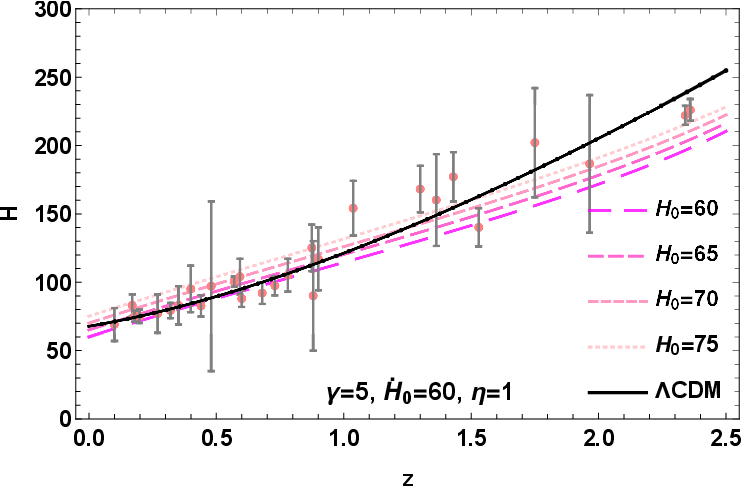}\hspace{2mm}
\includegraphics[width=.45\textwidth]{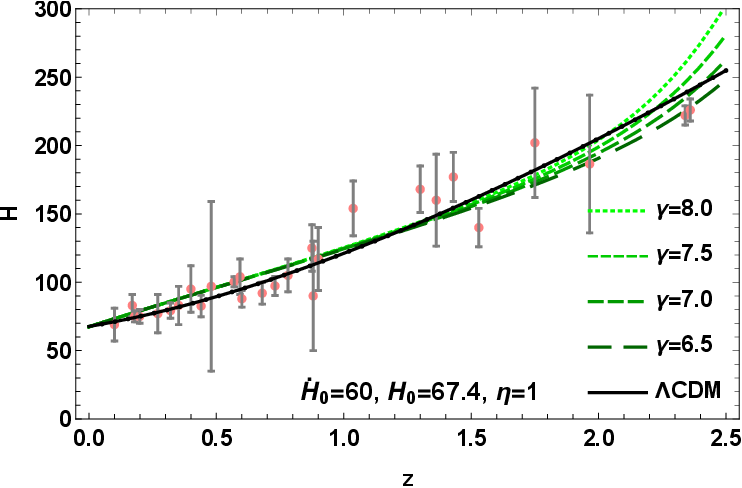}\vspace{2mm}
\includegraphics[width=.45\textwidth]{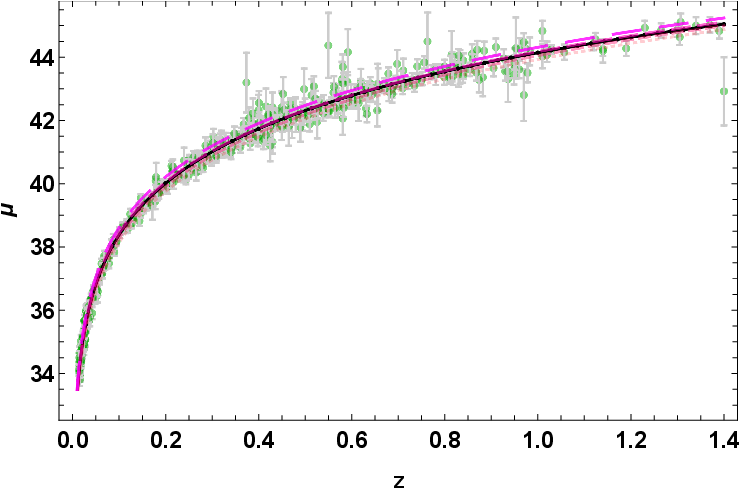}\hspace{2mm}
\includegraphics[width=.45\textwidth]{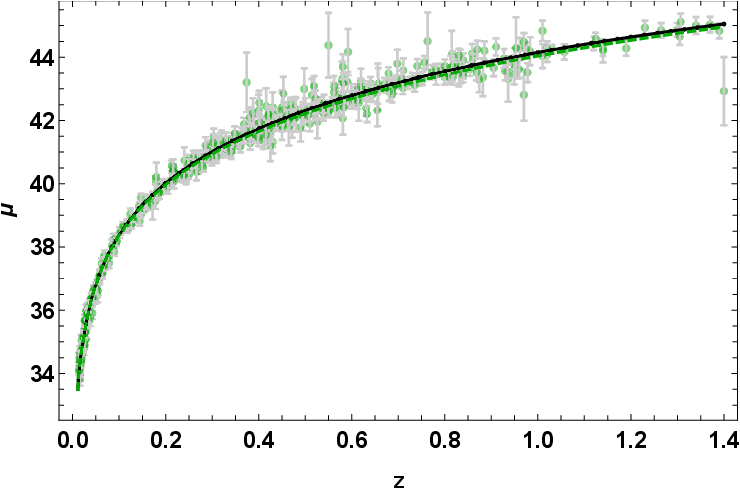}
\caption{The Hubble parameter and the distance modulus resulted from $f(Q)=\eta Q^{\frac{3}{2}}$ gravity. The lower panels follow the upper ones model parameters. $a_{0}=1=\rho_{0}$ has been set.}\label{noncoin.H1}
\end{figure}

Now returning eqs.~(\ref{noncoinH1})-(\ref{noncoinH2}) to the redshift space owing to the relations $a(z)=a_{0}(1+z)^{-1}$, $H'(t)=-(z+1) H(z) H'(z)$ and $H''(t)=(z+1) H(z) [(z+1) H(z) H''(z)+(z+1) H'(z)^2+H(z) H'(z)]$ we plot the behavior of the Hubble parameter and also the distance modulus in Fig.~\ref{noncoin.H1}. The left panels of Fig.~\ref{noncoin.H1} have been drawn for $\eta=1$, $\gamma=5$, $\dot{H}_{0}=60$ and different values of the Hubble parameter while for the right panels we applied different values of $\gamma$, $\eta=1$, $H_{0} = 67.4~\textrm{km s}^{-1} \textrm{Mpc}^{-1}$~\cite{planck2018} and the same value of  $\dot{H}_0$. Astronomical data reported in~\cite{amanullah2010} were used for the lower panels of Fig.~\ref{noncoin.H1}. In all panels the black solid curve denotes the $\Lambda CDM$ ones. 

\begin{figure}[h!]
\centering
\includegraphics[width=.45\textwidth]{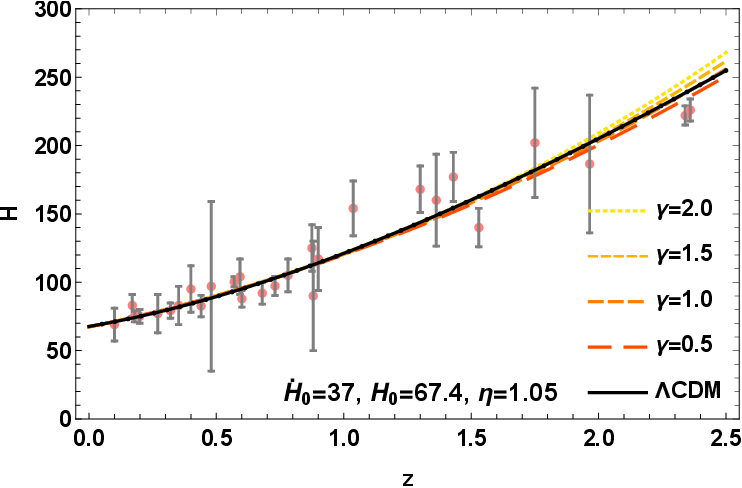}
\includegraphics[width=.45\textwidth]{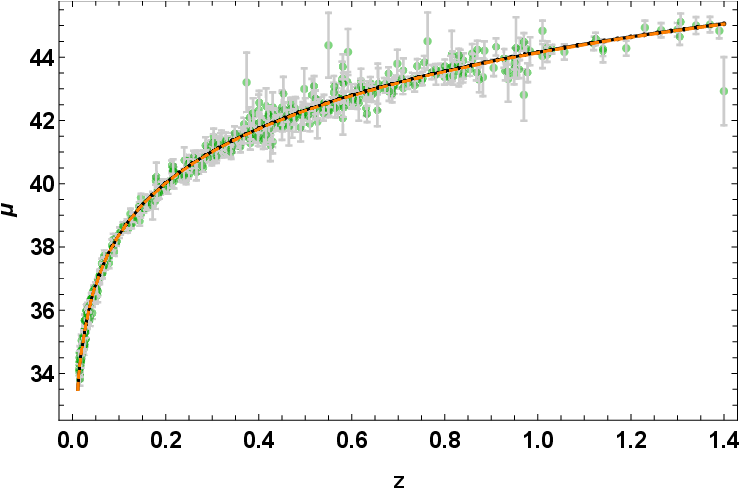}
\caption{The Hubble parameter and the distance modulus depicted for $f(Q)=1.05~Q^{0.9}$ gravity}\label{noncoin.H2}
\end{figure}

It is interesting to see the effects of other values of the power $n$ in $f(Q)=\eta Q^{n}$. In this case, eqs.~(\ref{noncoinH3})-(\ref{noncoinH4}) are calculated using eqs.~(\ref{noncoin.1})-(\ref{noncoin.3}). We have illustrated the diagrams of the Hubble parameter and the distance modulus for $n=0.9$, $\eta=1.05$, $0.5<\gamma<2.0$, $H_0=67.4$ and $\dot{H}_{0}=37$ in Fig.~\ref{noncoin.H2}. As can be seen, in this case a better consistency to the $\Lambda CDM$ curves is accessed.

\begin{align}
&6^{n-1} (n-1) n \eta H^{2 n-4} \Bigg\{H H'' \left(4 a^3 H-3 \gamma \right)+H' \Big[a^3 \left(6 H^3+H\right)+(2 n-3) H' \left(4 a^3 H-3 \gamma \right)\Big]\Bigg\}\notag\\
&+a^3\left(3 H^2+2 H'\right)=0,\label{noncoinH3}\\
&\eta\gamma  2^{n-1} 3^n (n-1) n H^{2 n-4} \left[H H''+(2 n-3) \left(H'\right)^2+3 H^2 H'\right]-a^3\left(\dot{\rho}+3H\rho_{m}\right)=0.\label{noncoinH4}
\end{align}

\subsection{Including ultra-relativistic matter}\label{noncoin1s2}
\begin{figure}[h!]
\centering
\includegraphics[width=.45\textwidth]{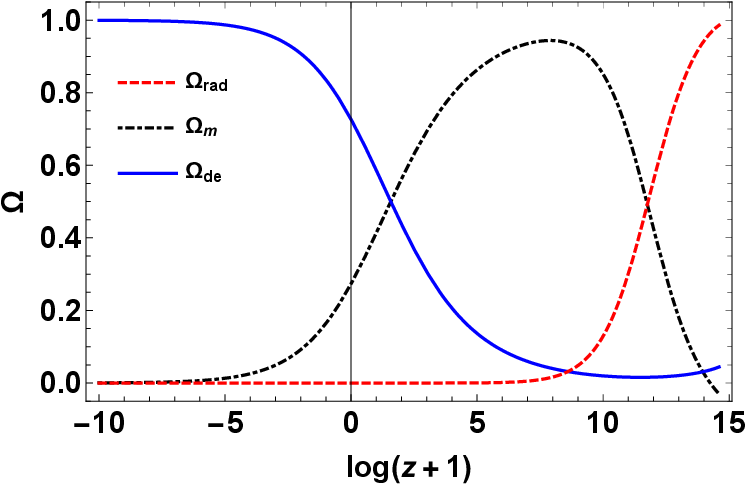}\hspace{2mm}
\includegraphics[width=.45\textwidth]{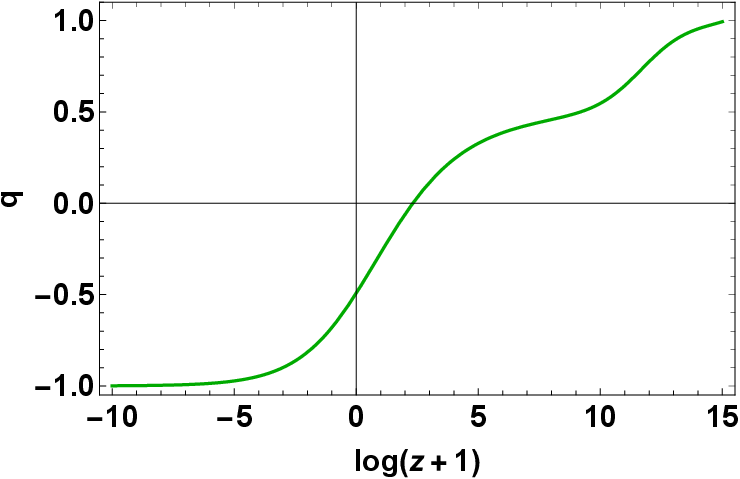}
\caption{Cosmological parameters in $f(Q,C)=Q^{\frac{5}{4}}+\alpha C^{\frac{4}{4}}$ gravity. Left panel: matter density parameters; the red dashed curve denotes the evolution of the ultra-relativistic matter. Right panel: the deceleration parameter. We have used the initial values $x_{1i}=4.92 \times 10^{-2}$, $x_{3i}=-1.99$, $x_{4i}=-1\times 10^{-2}$ for $\alpha=-3.1$ and $\gamma=-1.3$.}\label{noncoin13}
\end{figure}

In the above study only the pressureless perfect fluid has been considered. By contributing the ultra-relativistic perfect fluid some equations change, as follows

\begin{align}
&\Omega_{m}=1-\Omega_{de}-\Omega_{rad},\label{noncoin.14}\\
&x_{2}=\frac{2 \left(x_3+3\right)^2 \Big\{(n-1) x_1 \big[n x_3 \left(3 \gamma  x_4-4\right)-3\big]+2 x_3+3+\Omega_{rad}\Big\}}{9 \alpha  \gamma  m(m-1) x_3 x_4 \big[(2 n-3) x_3-3\big]+6 \alpha  (m-1) \left(x_3+3\right)^2+6 x_3 \big[(3-2 n) x_3+3\big]},\label{noncoin.15}
\end{align}

and the evolutionary equation for $\Omega_{rad}$

\begin{align}\label{noncoin.16}
\frac{d\Omega_{rad}}{dN}=-2\Omega_{rad} (2 + x_{3}).
\end{align}

is added to the system (\ref{noncoin.11})-(\ref{noncoin.13}). The attractor de Sitter solution is still present with eigenvalues $(0,-3,-3,-4)$, however, a saddle solution (with the eigenvalues $(-1,-1,0,3)$) indicating the radiation dominated era exists provided $n=5/4, m=5/4$. This tells us that different matter dominated eras play the role for different values of $n$ and $m$. Nevertheless, as Fig.~\ref{noncoin13} exhibits, a proper transition between different cosmological era can take place. The Universe evolves from a state dominating the radiation matter (for which one has $q=1$), temporarily stays in a situation in which the dark matter dominates (with $q=0.5$) and finally acceleratingly expand for ever (tending to $q=-1$).

\section{Connection type $III$}\label{noncoin1.2}
In the case of connections $\Gamma_{III}$ the following set of equations are obtained \cite{als/fQC}

\begin{align}
&\kappa \rho_m=3H^2+\frac{f}{2}+3H^{2}f_{Q}-C\frac{f_{C}}{2}-\frac{3\gamma}{2a^{3}}\dot{f}_{Q}+\left(3H+\frac{3\gamma}{2a^{3}}\right)\dot{f}_{C},\label{noncoinn.1}\\
&\kappa p_m=-2\dot{H}-3H^{2}-\frac{f}{2}-3H^{2}f_{Q}+C\frac{f_{C}}{2}+\left(\frac{\gamma}{2a^{3}}-2H\right)\dot{f}_{Q}-\frac{\gamma}{2a^{3}}\dot{f}_{C}-\ddot{f}_{C},\label{noncoinn.2}\\
&\dot{\rho}_{m}+3H(\rho_{m}+p_{m})=\frac{3\gamma}{2\kappa a^{3}}\left[H\dot{f}_{Q}-\ddot{f}_{Q}-(H\dot{f}_{C}-\ddot{f}_{C})\right].\label{noncoinn.3}
\end{align}

Rewriting eqs.~(\ref{noncoinn.1})-(\ref{noncoinn.3}) in terms of  the variables (\ref{noncoin.6}), assuming $f(Q,C)=Q^{i}+\beta C^{j}$\footnote{In this section, we use $i$ instead of $n$, $j$ instead of $m$ and $\beta$ for $\alpha$ to avoid any confusion.} and using

\begin{align}
&H\dot{f}_{Q}-\ddot{f}_{Q}=0,\label{noncoinn.4}\\
&H\dot{f}_{C}-\ddot{f}_{C}=0,\label{noncoinn.5}
\end{align}

one gets the same equations as (\ref{noncoin.9}) and (\ref{noncoin.11})-(\ref{noncoin.13}) achieving new equations for $x_2$, $x_5$ and $\Omega_{de}$ which are

\begin{align}
&x_{2}=\frac{2 \left(x_3+3\right)^2 \Big\{(i-1) x_1 \left[i x_3 \big(3 \gamma  x_4-8\big)-6\right]+4 x_3+6\Big\}}{3 \beta  \gamma  (j-1) j x_3 x_4 \left[(2 i-1) x_3-7\right]+4 x_3 \left[(2 i-1) x_3-7\right]+12 \beta  (j-1) \left(x_3+3\right){}^2},\label{noncoinn.6}\\
&x_{5}=x_3\frac{x_3(1-2 i )+7}{6 \left(x_3+3\right)^2}.\label{noncoinn.7}\\
&\Omega_{de}=(i-1) x_1 \big[1-\gamma  i x_3 x_4\big]+\beta  (j-1) x_2 \big[1-3 j x_5 \left(\gamma  x_4+2\right)\big].\label{noncoinn.8}
\end{align}

\begin{figure}[h!]
\centering
\includegraphics[width=.45\textwidth]{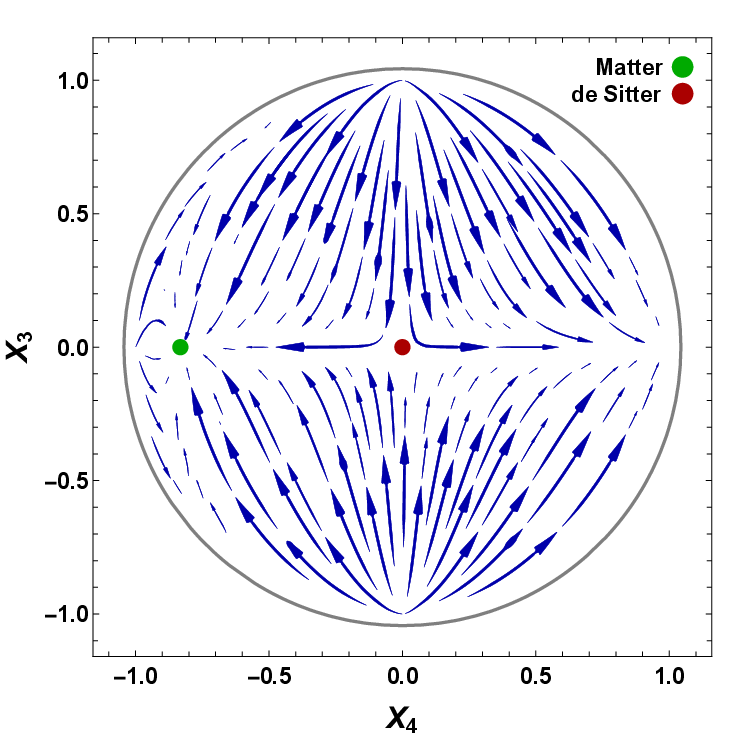}\vspace{2mm}\\
\includegraphics[width=.45\textwidth]{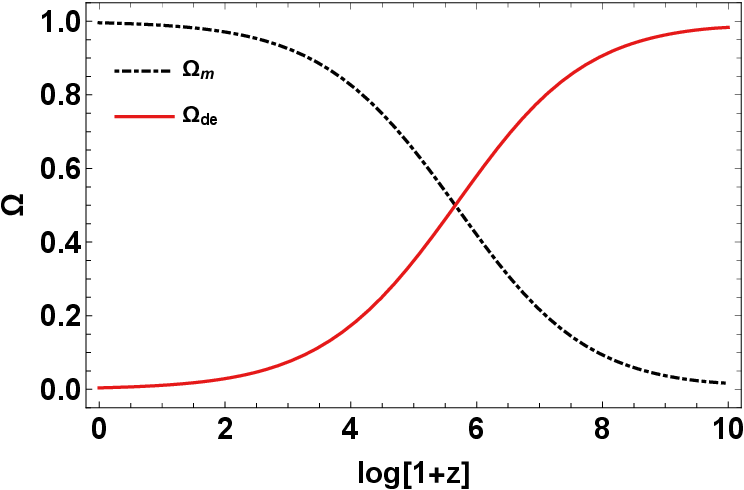}\hspace{2mm}
\includegraphics[width=.45\textwidth]{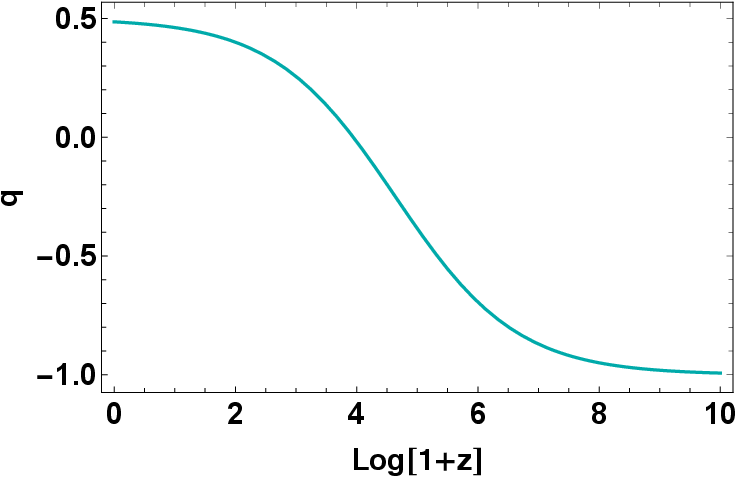}
\caption{Upper panel: Phase space portrait provided for the models $f(Q,C)=Q^{-\frac{5}{6}}+\alpha C^{\frac{7}{9}}$ including thee connection type $\Gamma_{III}$. Lower panels: the brown solid curve and the black dashed one (the left panel) indicate the dark energy and the dark matter densities, respectively. Also, the cyan curve (the right panel) displays the behavior of the deceleration parameter. }\label{noncoin14}
\end{figure}

Table~\ref{tab1} shows the critical points of the system including eqs.~(\ref{noncoin.11})-(\ref{noncoin.13}) with (\ref{noncoinn.6})-(\ref{noncoinn.8}). These equations accept a saddle de Sitter fixed point as well as a saddle dark matter one with $\Omega_m=1$ and $w_{eff}=0$ for $i= -\frac{5}{6} ,j=\frac{7}{9}$. In fact, in the case of the connection type $\Gamma_{III}$, the de Sitter solution corresponds to an early time accelerated expansion of the Universe. The behavior of the phase space trajectories in the plane $(x_4,x_3)$ illustrated in Fig.~\ref{noncoin14}. A transition from the de Sitter to the dark matter dominated eras is understood from Fig.~\ref{noncoin14}.

\section{Concluding remarks}\label{sec5}

The introduction of the boundary term $C$ in the non-metricity based gravity theories not only lifts the second-order $f(Q)$ theory into fourth order and generalises the $f(\mathring R)$ theory as a subclass, it also gives a rather much comprehensive phenomenology. In the present study, we have attempted to give a complete cosmological picture under the $f(Q,C)$ theories of gravity. First, we have proposed a number of variables and dimensionless parameters in each consecutive case to formulate the interconnecting equations for the dynamical system approach. The conservation of the energy-momentum tensor results in a number of constraint equations (see eqs.~(\ref{noncoin.4})-(\ref{noncoin.5}) and~(\ref{noncoinn.4})-(\ref{noncoinn.5})). We explored the possible three classes of formulations viable in this theory~\cite{FLRW/connection, ad/bianchi, fQec1} in spatially flat FLRW spacetime geometry:

\begin{itemize}

\item[I)] which uses connection type $I$ and gives eqs.~(\ref{coin.1})-(\ref{coin.2}). Two classes of models have been studied.
	
\begin{itemize}
\item[*] models with $f(Q,C)=Q+dC^{2}$ which include critical points corresponding to the dark matter, de Sitter, stiff fluid and phantom eras. Hence, different scenarios can be understood within the model. An interesting case is 		          transition from the de Sitter era to the dark matter era followed by a phantom dark energy dominated epoch (see the plot     	of the decelerated parameter depicted in the middle right panel of Fig.~\ref{coin1}.). The Universe experiences 		           	accelerated expansion both in the early and the late times. The de Sitter solution is unstable which is accounted for the 			beginning stage of the evolution of the Universe. Also, the accelerated expansion in the late times is derived by a 		           phantom field with $q=-2$ which ia mathematically an attractor. A double de Sitter solution (they act as the early and the late times eras) appears in $f(Q)$ gravity (in a spatially flat FRW geometry) using a non-trivial exponential function~\cite{shabani2023}. Nevertheless, here, we extract similar solution  by adding a boundary term to the trivial function $f(Q)=Q$. 
	
	\item[*] models with $f(Q,C)=Q^{s}+eC^{r}$ which again accept the mentioned critical points. Here, the de Sitter 			solution is stable for $1.00<s\lesssim1.16$ while the phantom dark energy solution is always unstable. Phase transition 			between the dark matter era to the dark energy one with the de Sitter era as the last stage of the evoultion of the 			Universe can be conceived.
	\end{itemize}

\item[II)] the connection type $II$ is used in this case for which  eqs.~(\ref{noncoin.1})-(\ref{noncoin.1}) are obtained. Gravitational models which can be described by the function $f(Q,C)=Q^{n}+\alpha C^{m}$ have been studied. These models explain a cosmic phase transition from a matter dominated to a dark energy dominated epoch for which a de Sitter solution is responsible. Also, by adding the contribution to the ultra-relativistic matter to equations we achieved a true sequences of radiation $\to$ dark matter $\to$ dark energy (see panels in Fig.~\ref{noncoin13}). In addition, we have drawn the related diagrams of the Hubble and the distance modules for $f(Q,C)=\eta Q^{n}$. We found more consistency to astronomical data for $n=0.9$ and $\eta=1.05$.

\item[III)] finally, we have theories which employ the connection type $III$ and lead to eqs.~(\ref{noncoinn.1})-(\ref{noncoinn.3}). The same $f(Q,C)$ function as that of item II was investigated. Here, a transition from an early accelerated expansion which is described by an unstable de Sitter solution to the dark matter era is perceived

\end{itemize}

Summarising our findings, there are both unstable and stable de-Sitter solutions that correlate to accelerated expansions in the early and late stages. We have clearly demonstrated all the findings in Table \ref{tab0} -- Table \ref{tab1}; the phase-space diagrams are provided.

\end{document}